\documentclass[aps,prd,onecolumn,groupedaddress,showpacs,nofootinbib,amssymb]{revtex4}
\usepackage{graphicx,bm,color}
\usepackage{amsmath}
\usepackage{amssymb}
\usepackage{amsfonts}
\usepackage{wrapfig}
\usepackage{cancel}

\newcommand{\be}{\begin{equation}}
\newcommand{\ee}{\end{equation}}
\newcommand{\bea}{\begin{eqnarray}}
\newcommand{\eea}{\end{eqnarray}}
\newcommand{\beaa}{\begin{eqnarray*}}
\newcommand{\eeaa}{\end{eqnarray*}}

\newcommand{\nn}{\nonumber \\}
\newcommand{\e}{\mathrm{e}}
\newcommand{\tr}{\mathrm{tr}\,}

\newcommand{\bk}{\bm{k}}
\newcommand{\bl}{\bm{l}}
\newcommand{\bx}{\bm{x}}

\newcommand{\bn}{\bm{n}}
\newcommand{\bmm}{\bm{m}}

\allowdisplaybreaks[4]

\begin{document}

\title{On the propagation of gravitational waves in strong  magnetic fields}

\author{Kazuharu Bamba$^{1}$, Shin'ichi Nojiri$^{2,3,4}$ and
Sergei D. Odintsov$^{5,6,7}$
}
\affiliation{
$^1$Division of Human Support System, Faculty of Symbiotic
Systems Science, Fukushima University, Fukushima 960-1296, Japan\\
$^2$Department of Physics, Nagoya University, Nagoya 464-8602, Japan\\
$^3$Kobayashi-Maskawa Institute for the Origin of Particles and the
Universe, Nagoya University, Nagoya 464-8602, Japan\\
$^4$ KEK Theory Center, High Energy Accelerator Research Organization (KEK),
Oho 1-1, Tsukuba, Ibaraki 305-0801, Japan \\
$^5$Instituci\'{o} Catalana de Recerca i Estudis Avan\c{c}ats
(ICREA), Passeig Llu\'{i}s Companys, 23, 08010 Barcelona, Spain\\
$^6$ Institute of Space Sciences (ICE, CSIC) C. Can Magrans s/n, 08193
Barcelona, Spain \\
$^7$ Institute of Space Sciences of Catalonia (IEEC), Barcelona, Spain
}

\begin{abstract}
The propagation of gravitational waves is explored in the cosmological context.
It is explicitly demonstrated that the propagation of gravitational waves could
be influenced
by the medium. It is shown that in the thermal radiation, the propagation
of gravitational waves in general relativity is different from that in the
scalar-tensor theory.
The propagation of gravitational waves is investigated in the uniform magnetic
field.
As a result, it is found that cosmic magnetic fields could influence on the
propagation of
gravitational waves to non-negligible extent. The corresponding estimation
for the spiral galaxy NGC 6946 effect is made.
\end{abstract}

\hspace{14.5cm} FU-PCG-28

\maketitle

\section{Introduction}

It has been proved by LIGO that two-black-holes system emits strong
gravitational waves in the coalescence phase~\cite{Abbott:2016blz}.
The first detection was from black holes with about 30 solar masses and
the following ones were from the mergers of two 
black holes (black hole binary)~\cite{Abbott:2016nmj, Abbott:2017vtc, Abbott:2017oio,
Abbott:2017gyy}.
Very recently, the so-called multi messenger astronomy has started by
the discovery of strong gravitational waves from the collision of
two neutron stars~\cite{TheLIGOScientific:2017qsa} 
and the electromagnetic radiation was detected in coincidence with the gravitational wave. 

It is very difficult and complicated to analyze the processes of
black hole mergers and scatterings because gravitational dynamics
is too strong. 
In spite of the difficulties, there has been accurate numerical simulations, 
which reproduce the observational results
\cite{Campanelli:2005dd,Mroue:2013xna}
although various approximate
approaches~\cite{Blanchet:2013haa}
and analytic ideas~\cite{Camps:2017gxz, McCarthy:2018zze} to calculate
the gravitational wave signatures in the strong gravitational field regimes
have been also proposed.

On the other hand, the existence of cosmic magnetic fields have been known and
those origins have also been explored.
In particular, the origins of large-scale magnetic fields
observed in clusters of galaxies can be primordial magnetic fields
from inflation and the following cosmological phases in the early universe
(for reviews on cosmic magnetic fields, see, e.g.,~\cite{Kronberg:1993vk,
Grasso:2000wj,Carilli:2001hj, Widrow:2002ud, Giovannini:2003yn, Giovannini:2004rj,
Giovannini:2006kg, Kandus:2010nw, Yamazaki:2012pg, Durrer:2013pga}).

Moreover, various modified gravity theories including
the scalar-tensor theory have especially been studied
in the cosmological context recently in order to explain the late-time cosmic
acceleration
(for recent reviews on modified gravity theories as well as dark energy
problem, see,
for example,~\cite{Nojiri:2010wj, Capozziello:2011et, Nojiri:2017ncd,
Capozziello:2010zz, Bamba:2012cp}).
The cosmological bounds from the Neutron Star Merger
GW170817~\cite{{TheLIGOScientific:2017qsa}} have been examined in the
scalar-tensor and $F(R)$ gravity theories~\cite{Nojiri:2017hai}.
The constraints~\cite{DeLaurentis:2016jfs} on alternative theories of gravity
have been calculated with GW150914 and
GW151226~\cite{TheLIGOScientific:2016src, Abbott:2016nmj,
TheLIGOScientific:2016pea}.
Various features of gravitational waves from modified gravity theories
have also been studied~\cite{{Capozziello:2008rq, Bellucci:2008jt,
Bogdanos:2009tn}}.

In this paper, we clarify how the propagation of gravitational waves could be
changed by the medium.
Usually the radiation is made of the quanta or relativistic particles at the high
temperature as in the early universe after the inflation.
In the radiation dominated era, the universe expandes as $a \propto t^{-\frac{1}{2}}$.
Here $a$ is the scale factor of the universe and $t$ is the cosmological time. 
On the other hand, it is known that the power law behavior
$a \propto t^{-\frac{1}{2}}$ in the radiation dominated universe can be
also realized by the classical scalar-tensor theory.\footnote{
For example, it has been shown that any evolution of the universe expansion 
can be realized in the scalar-tensor theory in \cite{Capozziello:2005tf}. 
See also \cite{Fujii:2003pa,Faraoni:2004pi}. 
}
In order to distinguish the above two kinds of the radiation dominated
universe, we show that the propagation of gravitational waves
in the thermal radiation in general relativity is different from that in the
classical scalar-tensor theory.
Usually, the radiation is made of photons, which are quanta of the
electromagnetic field.
The classical electromagnetic field is different from the photon.
As an example of the classical electromagnetic field,
we investigate the propagation of gravitational waves in the uniform magnetic
field and
we demonstrate that the effects from the magnetic field to the propagation
could not be negligible.

The structure of the paper is the following.
In Sec.~\ref{SecI}, we explore the propagation of gravitational waves in
general matter.
In Sec.~\ref{SecII}, we investigate the propagation in quanta and thermal
matter with finite temperature.
In Sec.~\ref{SecIII}, we analyze the propagation of gravitational waves under
the existence of magnetic fields.
In Sec.~\ref{SecIV}, we consider the case of gravitational waves in $F(R)$
gravity.
Finally, conclusions are given in Sec.~\ref{SecV}.

\section{Propagation of Gravitational Wave in Matters \label{SecI}}

The gravitational wave is given by the perturbation from the background
geometry,
\begin{equation}
\label{H1}
g_{\mu\nu} \to g_{\mu\nu} + \kappa^2 h_{\mu\nu}\, ,
\end{equation}
where $|h_{\mu\nu}|\ll 1$ is the perturbation with respect to a given
background $g_{\mu\nu}$.
Then by imposing the gauge condition
\begin{equation}
\label{H4}
\nabla^\mu h_{\mu\nu} = g^{\mu\nu} h_{\mu\nu} = 0\, ,
\end{equation}
the Einstein field equations
\begin{equation}
\label{einstein}
R_{\mu\nu} - \frac{1}{2} g_{\mu\nu} R = \kappa^2 T_{\mu\nu}\, ,
\end{equation}
take the perturbed form as follows,
\begin{equation}
\label{H6}
\frac{1}{2} \left[ - \nabla^2 h_{\mu\nu}
   - 2 R^{\lambda\ \rho}_{\ \nu\ \mu} h_{\lambda\rho}
+ R^\rho_{\ \mu} h_{\rho\nu} + R^\rho_{\ \nu} h_{\rho\mu}
    - h_{\mu\nu} R + g_{\mu\nu} R^{\rho\lambda} h_{\rho\lambda} \right]
= \kappa^2 \delta T_{\mu\nu}\, .
\end{equation}
Let denote the scale of $\kappa^2 T_{\mu\nu}$ by $M^2$, that is,  
$\kappa^2 T_{\mu\nu} \sim M^2$. 
If we assume that $M^2$ can be small enough and we can expand the 
l.h.s. and the r.h.s. of the Einstein equation with respect to $M^2$ as 
\begin{equation}
\label{P1}
R_{\mu\nu} - \frac{1}{2} g_{\mu\nu} R =I_R^{(0)} + M^2 I_R^{(1)} + M^4 I_R^{(2)} 
+ \mathcal{O}\left( \left( M^2 \right)^3 \right) \, , \quad 
\kappa^2 T_{\mu\nu}= M^2 I_T^{(1)} + M^4 I_T^{(2)} 
+ \mathcal{O}\left( \left( M^2 \right)^3 \right) \, .
\end{equation}
We should note that the r.h.s. starts with $\mathcal{O}\left( M^2 \right)$ term 
and therefore the $\mathcal{O}\left( 1 \right)$ term $I_R^{(0)}$ in the l.h.s. 
should vanish, which gives flat vacuum solution $g_{\mu\nu}=\eta_{\mu\nu}$. 
Then $\mathcal{O}\left( M^2 \right)$ term $M^2 I_T^{(1)}$, which expresses the 
matters in the flat background and equation $M^2 I_R^{(1)} = M^2 I_T^{(1)}$ gives 
the $\mathcal{O}\left( M^2 \right)$ correction to the geometry. 
We should note that the energy-momentum tensor $T_{\mu\nu}$ in 
Eq.(\ref{einstein}) depends on the metric, therefore the 
$\mathcal{O}\left( M^4 \right)$ term $M^4 I_T^{(2)} $ in the r.h.s. expresses 
the matter in the background with the $\mathcal{O}\left( M^2 \right)$ correction. 
Then the equation $M^4 I_R^{(2)}  = M^4 I_T^{(2)} $ gives the 
$\mathcal{O}\left( M^4 \right)$ correction to the geometry. 
By iterating the above procedure, we can find the background geometry by the perturbation 
with respect to $M^2$. 
The corrections to the geometry, which includes the gravitational 
wave, appears as the perturbative series with respect to $\kappa^2$. 
Therefore we have two parameters $M^2$ and $\kappa^2$ for the perturbative 
expansions. 
The parameter $M^2$ is conceptually different from the parameter $\kappa^2$ and 
they are independent with each other. 
Then the l.h.s. and the r.h.s. of the Einstein equation can be expressed by the 
double expansion with respect to $M^2$ and $\kappa^2$ 
\begin{align}
\label{P2}
R_{\mu\nu} - \frac{1}{2} g_{\mu\nu} R =& I_R^{(0)} + M^2 I_R^{(1,0)} + \kappa^2 I_R^{(0,1)} 
+ M^2 \kappa^2 I_R^{(2)} + \mathcal{O}\left( M^4, \kappa^2, \kappa^2 M^2 \right) \, , \nn
\kappa^2 T_{\mu\nu}=& M^2 I_T^{(1,0)} + M^2 \kappa^2 I_T^{(1,1)} 
 + \mathcal{O}\left( M^4, \kappa^4 \right) \, .
\end{align}
In Eq.~(\ref{P2}), $\mathcal{O}\left( \kappa^2 \right)$, $\kappa^2 I_R^{(0,1)}$, 
$M^2 \kappa^2 I_R^{(2)}$, $M^2 \kappa^2 I_T^{(1,1)}$, and 
$\mathcal{O}\left( \kappa^2 M^4 \right)$ terms expresses the 
propagation of the gravitational wave because the energy-momentum tensor 
$T_{\mu\nu}$ in Eq.~(\ref{einstein}) depends on the metric and if we consider 
the perturbation as (\ref{H1}), there is a variation of the energy-momentum tensor 
$T_{\mu\nu}$ in (\ref{H6}). 
On the other hand, the $\mathcal{O}\left( \kappa^4 \right)$ terms 
includes the non-linear interaction between the gravitational wave. 
We now neglect the interactions between the gravitational wave, we omit the 
$\mathcal{O}\left( M^4, \kappa^4 \right)$ terms and consider the 
$\kappa^2$ terms including the leading corrections 
with respect to $M^2$, that is, $\kappa^2 I_R^{(0,1)}$, 
$M^2 \kappa^2 I_R^{(2)}$, and $M^2 \kappa^2 I_T^{(1,1)}$.
The effect of the term $M^2 \kappa^2 I_T^{(1,1)}$ 
is similar to the propagation of light in the medium
(like water).
As we know by the Cerenkov radiation, the speed of the light decreases in the medium.
Although the propagation of the light follows also the  geodesics, the incident light makes
the electric charge or electric or magnetic moment distributions fluctuate and 
the fluctuations with the electric or magnetic dipole moments generate the light.
The generated light interferes with the incident light and the decrease of the propagation
speed of the light occurs.
This effects are known as a polarization and can be expressed as the changes 
of the permittivity and permeability.
Even for the gravitational wave, there occurs similar phenomena, which was also recently
reported in the paper by Weinberg et~al. \cite{Flauger:2017ged} in detail 
for the propagation of the gravitational wave in the cold dark matter.
The incident gravitational wave makes the medium fluctuate and the fluctuation with
quadrupole moment generates an additional gravitational wave.
The r.h.s. in (\ref{H6}) or the term $M^2 \kappa^2 I_T^{(1,1)}$ in (\ref{P2}) 
expresses such effects although our formulation is rather simplified
compared with the paper \cite{Flauger:2017ged}.

\section{Quanta and Thermal Matter  \label{SecII}}

In this section, we consider the real scalar field as the matter.
We treat the scalar field as the quantum field at the finite temperature.
In case of the high temperature or in massless case, the scalar field
plays the role of the radiation.
On the other hand, in the limit that the temperature vanishes but the density
is finite, we obtain the dust, which can be a cold dark matter.
After that, we compare the obtained results with those in the classical
scalar-tensor model~\cite{Capozziello:2008fn, Nojiri:2017hai}.

Even in the classical scalar-tensor model, we can realize matter dominated
(filled with dust) or radiation dominated universe.
Then we find that in case of the matter dominated universe, the result
for the quantum field coincides
with the result in the classical scalar-tensor theory but in other case, the
tensor structure
of $\delta T_{\mu\nu}$ is different in the two cases and therefore the
propagation of the gravitational wave changes in general.

In curved space-time, the energy-momentum tensor of a free real scalar
field $\phi$ with mass $M$ is given by
\begin{equation}
\label{Tphi1}
T_{\mu\nu} = \partial_\mu \phi \partial_\nu \phi
+ g_{\mu\nu} \left( - \frac{1}{2} g^{\rho\sigma} \partial_\rho \phi
\partial_\sigma \phi - \frac{1}{2} M^2 \phi^2 \right) \, .
\end{equation}
In the flat background, we find
\begin{align}
\label{Tphi3b}
T_{00} =& \rho = \frac{1}{2} \left( \pi^2 + \sum_{n=1,2,3}\left( \partial_n
\phi \right)^2
+ M^2 \phi^2 \right) \, , \nn
T_{ij} =& \partial_i \phi \partial_j \phi
+ \frac{1}{2} \delta_{ij} \left( \pi^2 - \sum_{n=1,2,3}
\left( \partial_n \phi  \right)^2 - M^2 \phi^2 \right) \, .
\end{align}
Here $\pi=\dot\phi$ is the momentum conjugate to $\phi$.
We also obtain
\begin{equation}
\label{Tphi2}
\frac{\partial T_{\mu\nu}}{\partial g_{\rho\sigma}}
= \frac{1}{2} \left( \delta_\mu^{\ \rho} \delta_{\nu}^{\ \sigma}
+ \delta_\mu^{\ \sigma} \delta_{\nu}^{\ \rho} \right)
\left( - \frac{1}{2} g^{\eta\zeta} \partial_\eta \phi \partial_\zeta \phi
   - \frac{1}{2} M^2 \phi^2 \right)
+ \frac{1}{2} g_{\mu\nu} \partial^\rho \phi \partial^\sigma \phi \, ,
\end{equation}
which has the following form in the flat background,
\begin{equation}
\label{Tphi3}
\frac{\partial T_{ij}}{\partial g_{kl}} = \frac{1}{4} \left( \delta_i^k
\delta_j^l + \delta_i^l \delta_j^k \right)
\left( \pi^2 - \sum_{n=1,2,3}\left( \partial_n \phi \right)^2 - M^2 \phi^2 \right)
+ \frac{1}{2} \delta_{ij} \partial^k \phi \partial^l \phi \, .
\end{equation}
We now evaluate $\frac{\partial T_{ij}}{\partial g_{kl}}$ in (\ref{Tphi3})
at
the finite temperature $T$.
In order to make the situation definite, we assume that the
three-dimensional
space is the square box where the lengths of the edges are $L$ and we impose
the periodic boundary condition on the scalar field $\phi$.
Then the momentum $\bk$ is given by
\begin{equation}
\label{box1}
\bk = \frac{2\pi}{L} \bn\, , \quad \bn =\left( n_x, n_y, n_z \right) \, .
\end{equation}
Here $n_x$, $n_y$, and $n_z$ are integers.
If we define,
\begin{equation}
\label{scalar8B}
\phi\left(\bx\right)
\equiv \frac{1}{L^\frac{3}{2}} \sum_{\bn}
\e^{i\frac{2\pi \bn \cdot \bx}{L}} \phi_{\bn} \, , \quad
\pi\left(\bx\right)
\equiv \frac{1}{L^\frac{3}{2}} \sum_{\bn}
\e^{i\frac{2\pi \bn \cdot \bx}{L}} \pi_{\bn} \, ,
\end{equation}
we find
\begin{equation}
\label{scalar9b}
\int d^3 x \phi\left(\bx\right)^2
= \sum_{\bn} \phi_{-\bn} \phi_{\bn}\, , \quad
\int d^3 x \pi\left(\bx\right)^2
= \sum_{\bn} \pi_{-\bn} \pi_{\bn}\, .
\end{equation}
The Hamiltonian is given by
\begin{equation}
\label{scalar1b}
H = \frac{1}{2} \sum_{\bn} \left( \pi_{-\bn} \pi_{\bn}
+ \left( \frac{4 \pi^2 \bn\cdot \bn}{L^2} + M^2 \right)
\phi_{-\bn} \phi_{\bn}\right) \, .
\end{equation}
Here $\tilde\pi(\bk)$ and $\tilde\phi(\bl)$ satisfy the following commutation
relation,
\begin{equation}
\label{scalar2b}
\left[ \pi_{\bn}, \phi_{\bn'} \right] = -i \delta_{\bn + \bn', 0} \, .
\end{equation}
We now define the creation and annihilation operators $a^\pm_{\bn}$ by
\begin{equation}
\label{scalar3b}
a^\pm_{\bn}
= \frac{1}{\sqrt{2}} \left( \frac{\pi_{\bn} }{\left( \frac{4 \pi^2 \bn\cdot
\bn}{L^2} + M^2 \right)^\frac{1}{4}}
\pm i \left( \frac{4 \pi^2 \bn\cdot \bn}{L^2} + M^2 \right)^\frac{1}{4}
\phi_{\bn} \right) \, .
\end{equation}
We should note $\left( a^\pm_{\bn} \right)^\dagger = a^\mp_{-\bn}$ because
${\pi_{\bn}}^\dagger = \pi_{-\bn}$ and ${\phi_{\bn}}^\dagger = \phi_{-\bn}$.
The operators $a^\pm_{\bn}$ satisfy the following commutation relation,
\begin{equation}
\label{scalar4b}
\left[ a^-_{\bn}, a^+_{\bn'} \right] = \delta_{\bn + \bn',0} \, ,
\quad \left[ a^\pm_{\bn}, a^\pm_{\bn'} \right] = 0 \, .
\end{equation}
The equations  (\ref{scalar3b}) can be solved with respect to $\pi_{\bn}$
and
$\phi_{\bn}$ as follows,
\begin{equation}
\label{scalar5b}
\phi_{\bn} = \frac{1}{i \left( \frac{4 \pi^2 \bn\cdot \bn}{L^2} + M^2
\right)^\frac{1}{4} \sqrt{2}} \left( a^+_{\bn} - a^-_{\bn}  \right) \, ,
\quad \pi_{\bn} = \frac{\left( \frac{4 \pi^2 \bn\cdot \bn}{L^2}
+ M^2 \right)^\frac{1}{4}}{\sqrt{2}}
\left( a^+_{\bn} + a^-_{\bn} \right) \, .
\end{equation}
The Hamiltonian (\ref{scalar1b}) can be rewritten as
\begin{equation}
\label{scalar6b}
H = \sum_{\bn} \sqrt{\frac{4 \pi^2 \bn\cdot \bn}{L^2} + M^2 }
\left(  a^+_{-\bn} a^-_{\bn} + \frac{1}{2} \right) \, .
\end{equation}
We now neglect the zero-point energy,
\begin{equation}
\label{scalar7b0}
H \to \tilde H = \sum_{\bn} \sqrt{\frac{4 \pi^2 \bn\cdot \bn}{L^2} + M^2 }
a^+_{-\bn} a^-_{\bn} \, .
\end{equation}
We define number operator by
\begin{equation}
\label{number}
N \equiv \sum_{\bn} a^+_{-\bn} a^-_{\bn} \, .
\end{equation}
Then we find the following expression of the partition function,
\begin{equation}
\label{scalar7b}
Z (\beta,\mu) = \tr \e^{ - \beta \tilde H - i \mu N }
= \e^{ - \sum_{\bn}
\ln \left(1 - \e^{ - \beta E_{\bn} - i\mu }\right)} \, ,
\quad E_{\bn} \equiv
\sqrt{ \frac{4 \pi^2 \bn\cdot \bn}{L^2} + M^2 } \, .
\end{equation}
Here $\beta=\frac{1}{k_\mathrm{B}T}$ with the Boltzmann constant
$k_\mathrm{B}$ and $\mu$ is the chemical potential.
Then we find the thermal average of the operator
$a^+_{\bmm} a^-_{\bn}$ is given as follows,
\begin{equation}
\left< a^+_{\bmm} a^-_{\bn} \right>_{T,\mu}
= - \delta_{\bmm + \bn,0}
\frac{1}{\beta} \frac{\partial \ln Z (\beta,\mu) }{\partial E_{\bn}}
= \delta_{\bmm + \bn,0}
\frac{\e^{ - \beta E_{\bn} - i\mu }}{1 - \e^{ - \beta E_{\bn} - i\mu }} \, .
\end{equation}
By normal ordering the operator $\frac{\partial T_{ij}}{\partial g_{kl}}$ in
(\ref{Tphi3}),
we acquire
\begin{align}
\label{scalar10D}
: \frac{\partial T_{ij}}{\partial g_{kl}} : =&
\frac{1}{L^3} \sum_{\bmm,\bn}
\e^{i\frac{2\pi \left(\bmm + \bn\right)\cdot \bx}{L}} \left\{
\frac{1}{4} \left( \delta_i^k \delta_j^l
+ \delta_i^l \delta_j^k \right)
\left( : \pi_{\bmm} \pi_{\bn}
+ \left( \frac{\left(2\pi\right)^2}{L^2} \bmm\cdot \bn - M^2 \right)
: \phi_{\bmm} \phi_{\bn} : \right) \right. \nn
& \qquad \left. - \frac{\left(2\pi\right)^2}{2L^2} \delta_{ij} m^k n^l
:\phi_{\bmm} \phi_{\bn} : \right\}  \nn
= & \frac{1}{L^3} \sum_{\bmm,\bn}
\e^{i\frac{2\pi \left(\bmm + \bn\right)\cdot \bx}{L}} \left\{
\frac{1}{4} \left( \delta_i^{\ k} \delta_j^{\ l} + \delta_i^{\ l} \delta_j^{\
k} \right)
\left( \sqrt{ E_{\bmm} E_{\bn} }
+ \frac{1}{\sqrt{ E_{\bmm} E_{\bn} }}
\left( \frac{\left(2\pi\right)^2}{L^2} \bmm\cdot \bn - M^2 \right)
\right) \right. \nn
& \qquad \left. - \frac{\left(2\pi\right)^2}
{2L^2 \sqrt{ E_{\bmm} E_{\bn} }} \delta_{ij} m^k n^l \right\}
a^+_{\bmm} a^-_{\bn} \, .
\end{align}
Therefore we obtain
\begin{equation}
\label{scalar11D}
\left< : \frac{\partial T_{ij}}{\partial g_{kl}} : \right>_T
= \frac{1}{2L^3} \sum_{\bn} \left\{
\frac{\left(2\pi\right)^2 n^k n^l }{L^2 E_{\bn}} \delta_{ij}
   \right\}
\frac{\e^{ - \beta E_{\bn} - i\mu }}{1 - \e^{ - \beta E_{\bn} - i\mu }}
= \frac{1}{6L^3} \sum_{\bn} \left\{
\frac{\left(2\pi\right)^2 \bn\cdot\bn }{L^2 E_{\bn}} \delta_{ij} \delta^{kl}
   \right\}
\frac{\e^{ - \beta E_{\bn} - i\mu }}{1 - \e^{ - \beta E_{\bn} - i\mu }}
\, .
\end{equation}
Particularly in case of massless, $M=0$, we find
\begin{equation}
\label{scalar12D}
\left< : \frac{\partial T_{ij}}{\partial g_{kl}} : \right>_{T,\, M=0}=
\frac{1}{6L^3}  \delta_{ij} \delta^{kl}
\sum_{\bn} \frac{2\pi \sqrt{\bn\cdot\bn}}{L}
\frac{\e^{ - \frac{2\pi \beta \sqrt{\bn\cdot\bn}}{L} - i\mu }}
{1 - \e^{ - \frac{2\pi \beta \sqrt{\bn\cdot\bn}}{L} - i\mu }}
\, .
\end{equation}
The expectation value of the number operator in (\ref{number}) is given by
\begin{equation}
\label{scalarDDD}
\left< N \right>_{T,\, M=0}=
\sum_{\bn} \frac{\e^{ - \beta E_{\bn} - i\mu }}
{1 - \e^{ - \beta E_{\bn} - i\mu }} \, .
\end{equation}

In the limit of $L\to \infty$, we obtain
\begin{align}
\label{scalar11}
\left< : \frac{\partial T_{ij}}{\partial g_{kl}} : \right>_T=&
\frac{1}{6 \left( 2\pi \right)^3} \delta_{ij} \delta^{kl}\int d^3 k
\frac{k^2}{\sqrt{k^2 + M^2}}
\frac{\e^{ - \beta \left( k^2 + M^2 \right)^\frac{1}{2} -i\mu }}
{1 - \e^{ - \beta \left( k^2 + M^2 \right)^\frac{1}{2} -i\mu }} \nn
=& \frac{1}{12\pi^2} \delta_{ij} \delta^{kl}\int_0^\infty dk
\frac{k^4}{\sqrt{k^2 + M^2}}
\frac{\e^{ - \beta \left( k^2 + M^2 \right)^\frac{1}{2} -i\mu }}
{1 - \e^{ - \beta \left( k^2 + M^2 \right)^\frac{1}{2} -i\mu }}
\, .
\end{align}
and in  massless case, $M=0$,
\begin{equation}
\label{scalar12}
\left< : \frac{\partial T_{ij}}{\partial g_{kl}} : \right>_{T,\, M=0}
= \frac{1}{12\pi^2}  \delta_{ij} \delta^{kl} \int_0^\infty dk
\frac{k^3 \e^{ - \beta k  -i\mu }} {1 - \e^{ - \beta  k -i\mu }} \, .
\end{equation}
The expectation value of the number density $n$ is given by
\begin{equation}
\label{ndensity}
\left< n \right>_{T,\, M=0}
\equiv \lim_{L\to\infty} \frac{\left< N \right>_{T,\, M=0}}{L^3}
= \frac{1}{2\pi^2} \int_0^\infty dk
\frac{k^2 \e^{ - \beta \left( k^2 + M^2 \right)^\frac{1}{2} -i\mu }}
{1 - \e^{ - \beta \left( k^2 + M^2 \right)^\frac{1}{2} -i\mu }} \, .
\end{equation}
By using (\ref{Tphi3b}), we also find
\begin{align}
\label{scalar14}
\left< \rho \right>_T=&  \frac{1}{2\pi^2} \int_0^\infty dk \frac{ k^2
\left( k^2 + M^2 \right)^\frac{1}{2}
\e^{ - \beta \left( k^2 + M^2 \right)^\frac{1}{2} -i\mu }}
{1 - \e^{ - \beta \left( k^2 + M^2 \right)^\frac{1}{2} -i\mu }} \, , \nn
\left< T_{ij} \right>_T = \delta_{ij} \left< p \right>_T =&
\frac{\delta_{ij}}{6 \pi^2 } \int_0^\infty d k \frac{ k^4
\e^{ - \beta \left( k^2 + M^2 \right)^\frac{1}{2} -i\mu }}
{\left( k^2 + M^2 \right)^\frac{1}{2}
\left( 1 - \e^{ - \beta \left( k^2 + M^2 \right)^\frac{1}{2} -i\mu }
\right)} \, .
\end{align}
In the massless limit $m\to 0$, we acquire
\begin{equation}
\label{scalar15-00}
\left< \rho \right>_T = 3 \left< p \right>_T = \frac{1}{4\pi^2} \int_0^\infty
dk \frac{
k^3 \e^{ - \beta k -i\mu }}
{1 - \e^{ - \beta k -i\mu }}
= \frac{1}{4\pi^2 \beta^4} \int_0^\infty ds
\frac{s^3 \e^{ - s -i\mu }}{1 - \e^{ - s -i\mu }} \, .
\end{equation}

When we explore the dark matter, the number of the particles might be
fixed.
Let the number be $N_0$, then the partition function in (\ref{scalar7b})
is replaced by
\begin{equation}
\label{scalar7bb}
Z_{N_0}(\beta) = \int_0^{2\pi} d\mu \e^{i \mu N_0} Z (\beta,\mu)
= \int_0^{2\pi} d\mu \e^{ i \mu N_0 - \sum_{\bn}
\ln \left(1 - \e^{ - \beta E_{\bn} - i\mu }\right)} \, .
\end{equation}
Especially if we consider the limit of $T\to 0$, only the ground state can
contribute and we find
\begin{equation}
\label{scalar11DN0}
\left< : \frac{\partial T_{ij}}{\partial g_{kl}} : \right>_{T=0, N=N_0} = 0\, ,
\end{equation}
and
\begin{equation}
\label{scalar14b}
\left< \rho \right>_{T=0, N=N_0} = \rho_0 \equiv \frac{N_0 M}{L^3}
\, , \quad
\left< T_{ij} \right>_{T=0, N=N_0} = 0\, .
\end{equation}

Until now, we have treated the scalar field as a quantum field at finite
temperature.
Instead of this, we often take the real scalar field as a classical field.
We now investigate if there is any difference in the two treatments.
The action of the general scalar field with potential has the following form:
\begin{equation}
\label{H7}
S_\phi = \int d^4 x \sqrt{-g} \mathcal{L}_\phi\, , \quad
\mathcal{L}_\phi =  - \frac{1}{2} \omega(\phi) \partial_\mu \phi
\partial^\mu \phi - V(\phi) \, .
\end{equation}
Then we find
\begin{equation}
\label{H8}
T_{\mu\nu} = - \omega(\phi) \partial_\mu \phi \partial_\nu \phi
+ g_{\mu\nu} \mathcal{L}_\phi\, ,
\end{equation}
and instead of (\ref{Tphi2}), we obtain
\begin{equation}
\label{Tphi2general}
\frac{\partial T_{\mu\nu}}{\partial g_{\rho\sigma}}
= \frac{1}{2} \left( \delta_\mu^{\ \rho} \delta_{\nu}^{\ \sigma}
+ \delta_\mu^{\ \sigma} \delta_{\nu}^{\ \rho} \right)
\left( - \frac{1}{2} g^{\eta\zeta} \omega(\phi) \partial_\eta \phi
\partial_\zeta \phi
   - V(\phi) \right)
+ \frac{1}{2} g_{\mu\nu} \omega(\phi) \partial^\rho \phi \partial^\sigma \phi
\, .
\end{equation}
When we assume the FRW universe with flat spatial part,
\begin{equation}
\label{FRWmetric}
ds^2 = - dt^2 + a(t)^2 \sum_{i=1,2,3} \left( dx^i \right)^2 \, ,
\end{equation}
and $\phi=t$ in (\ref{H7}),
a power-law behavior for the scale factor $a(t)$ of the universe,
\begin{equation}
\label{grv5}
a(t) = \left( \frac{t}{t_0} \right)^\alpha \, ,
\end{equation}
with  $t_0$ and $\alpha$ real constants,
can be realized by choosing
\begin{equation}
\label{LB2power}
\omega (\phi)  = \frac{2\alpha }{\kappa^2 t_0^2 \phi^2} \, , \quad
V(\phi) = \frac{3\alpha^2 - \alpha}{\kappa^2 t_0^2 \phi^2} \, .
\end{equation}
In case of the FRW universe (\ref{FRWmetric}) filled by the perfect fluid
whose equation of state (EoS) parameter $w$ is constant, $\alpha$ in
(\ref{grv5}) is given by
\begin{equation}
\label{GWp5}
\alpha = \frac{2}{3\left( 1 + w \right)} \, .
\end{equation}
For dust where $w=0$, in (\ref{Tphi2general}),
by using (\ref{LB2power})
and $\phi=t$, we find
\begin{equation}
\label{Gscalar1}
   - \frac{1}{2} g^{\eta\zeta} \omega(\phi) \partial_\eta \phi \partial_\zeta
\phi  - V(\phi) = \frac{1}{2} \omega(\phi) - V(\phi)
= \frac{2}{3 \kappa^2 t_0^2 \phi^2} - \frac{2}{3\kappa^2 t_0^2 \phi^2}
= 0 \, ,
\end{equation}
and therefore
\begin{equation}
\label{Tphi2general2}
\frac{\partial T_{ij}}{\partial g_{kl}} =0 \, ,
\end{equation}
which is consistent with (\ref{scalar11DN0}).
On the other hand, in case $w=\frac{1}{3}$, which corresponds to
the radiation, we obtain $\alpha=\frac{1}{2}$ and
\begin{equation}
\label{Tphi2general3}
\frac{\partial T_{ij}}{\partial g_{kl}}
= \frac{1}{2} \left( \delta_i^{\ k} \delta_{j}^{\ l}
+ \delta_i^{\ l} \delta_j^{\ k} \right)
\frac{1 }{4\kappa^2 t_0^2 \phi^2} \, ,
\end{equation}
whose tensor structure is different from that of the real radiation
in (\ref{scalar12}).

In general, in case of the quantum field at finite temperature,
we find the tensor structure of $\frac{\partial T_{ij}}{\partial g_{kl}}$ as
\begin{equation}
\label{scalar15}
\left< : \frac{\partial T_{ij}}{\partial g_{kl}} : \right>_T
\propto \delta_{ij} \delta^{kl}\, ,
\end{equation}
but for the tensor-scalar theory,
we find
\begin{equation}
\label{scalar16}
\frac{\partial T_{ij}}{\partial g_{kl}} \propto
\frac{1}{2} \left( \delta_i^{\ k} \delta_{j}^{\ l}
+ \delta_i^{\ l} \delta_j^{\ k} \right) \, .
\end{equation}
Due to the difference of the tensor structure, the propagation of the
gravitational wave
is different in the case of the quantum field at finite temperature and
the case
of the classical scalar-tensor theory, in general.
Especailly in case of the quantum field, because
$\left< : \frac{\partial T_{ij}}{\partial g_{kl}} : \right>_T$ always
includes
the factor $\delta^{kl}$,
by the condition $h_\mu^{\ \mu}=0$ in (\ref{H4}),
as long as we consider the gravitational
wave with $h_{tt}=0$, the term does not contribute.
We may investigate the radiation as a comprehensible example.
Usual radiation, for example in the early universe, is made of many quanta or
particles at finite temperature as is known in the (quantum) statistical
physics.
The radiation is realized by the massless particles or in the limit of
the high temperature.
On the other hand, the FRW universe in the radiation dominated era can be
realized by
the classical scalar-tensor theory.
The tensor structure of $\frac{\partial T_{ij}}{\partial g_{kl}}$ is, different
in the two cases,
as shown in (\ref{scalar15}) and (\ref{scalar16}).
The difference of the tensor structure generates the difference of the
propagation
of the gravitational wave~\cite{Capozziello:2008fn}.
In fact, the equation for the gravitational wave in the scalar-tensor theory is
given by
\begin{equation}
\label{LB3}
0=\left(2\dot H + 6 H^2 + H\partial_t-\partial_t^2
+\frac{\bigtriangleup}{a^2}\right)h_{ij}\, ,
\end{equation}
but in case of the quantum field with the finite temperature, we have
\begin{equation}
\label{LB3B}
0=\left(6\dot H + 12 H^2 + H \partial_t-\partial_t^2
+\frac{\bigtriangleup}{a^2}\right)h_{ij}\, ,
\end{equation}
where $\bigtriangleup$ is the Laplacian.

\section{Magnetic Field \label{SecIII}}

In this section, we analyze the propagation of the gravitational wave
under the magnetic field.
The energy-momentum tensor of the electromagnetic field in
   curved space-time is given by
\begin{equation}
\label{em1}
T_{\mu\nu} = g^{\rho\sigma} F_{\mu\rho} F_{\nu\sigma}
   -  \frac{1}{4} g_{\mu\nu} g^{\rho\sigma} g^{\eta\zeta} F_{\rho\eta}
F_{\sigma\zeta}\, , \quad
F_{\mu\nu} = \partial_\mu A_\nu - \partial_\nu A_\mu \, ,
\end{equation}
which gives
\begin{equation}
\label{em2}
\frac{\partial T_{\mu\nu} }{\partial g_{\rho\sigma}}
= - g^{\rho\eta} g^{\sigma\zeta} F_{\mu\eta} F_{\nu\zeta}
   - \frac{1}{8} \left( \delta_\mu^{\ \rho} \delta_\nu^{\ \sigma}
+ \delta_\mu^{\ \sigma} \delta_\nu^{\ \rho} \right)
   g^{\xi\tau} g^{\eta\zeta} F_{\xi\eta} F_{\tau\zeta}
+ \frac{1}{2} g_{\mu\nu} g^{\rho\xi} g^{\sigma\tau} g^{\eta\zeta}
F_{\xi\eta} F_{\tau\zeta} \, .
\end{equation}

Eq.~(\ref{H6}) shows that there are mainly two kinds of effects in the
magnetic
field.
The l.h.s. in (\ref{H6}) receives the change of the geometry
due to the existence of the magnetic field and we obtain non-trivial
connections and curvatures.
The r.h.s. tells that the gravitational wave gives some fluctuation of the
distribution of the
magnetic field, which becomes a new source of the gravitational field.

We should note that the contributions from the change of the geometry
are same order with the contributions from the fluctuation of the
magnetic field.

\subsection{Change of Geometry by Magnetic Field}

Because Eq.~(\ref{em1}) gives the effects via the fluctuation of the
distribution in the magnetic field, we now examine the change of the geometry.
We assume that the background is almost flat but there is a
constant magnetic field along the $z$ direction, $F_{xy}=-F_{yx}=B$.
Then we find
\begin{equation}
\label{GWM1}
T_{xx}=T_{yy} = \frac{1}{2}B^2 + \mathcal{O}\left( \kappa^2 \right)\, , \quad
T_{zz}= - T_{tt} = - \frac{1}{2}B^2 + \mathcal{O}\left( \kappa^2 \right)\, ,
\quad
\mbox{other components}=0 \, .
\end{equation}
The Einstein equation (\ref{einstein}) leads to
\begin{equation}
\label{curvatures}
R=0\, , \quad R_{\mu\nu} = \kappa^2 T_{\mu\nu} \, .
\end{equation}
The parameter $M^2$ in (\ref{P1}) and (\ref{P2}) corresponds to 
$\kappa^2 B^2$. 
Then before considering the gravitational wave, we need to
consider $\mathcal{O} \left( \kappa^2 B^2 \right)$ correction, 
corresponding to (\ref{P1}),  from the
flat background $g_{\mu\nu}=\eta_{\mu\nu}$,
\begin{equation}
\label{GWM2}
g_{\mu\nu} = \eta_{\mu\nu} + \kappa^2 B^2 \zeta_{\mu\nu}\, .
\end{equation}
Then the Einstein equation (\ref{einstein}) gives
\begin{equation}
\label{GWM3}
\partial_\mu \partial^\rho \zeta_{\nu\rho}
+ \partial_\nu \partial^\rho \zeta_{\mu\rho}
   - \partial_\rho \partial^\rho \zeta_{\mu\nu}
   - \partial_\mu \partial_\nu \left( \eta^{\rho\lambda} \zeta_{\rho\lambda}
\right)
   - \eta_{\mu\nu} \left( \partial^\rho \partial^\sigma \zeta_{\rho\sigma}
   - \partial^2 \left( \eta^{\rho\sigma} \zeta_{\rho\sigma}\right) \right)
= \frac{2}{B^2} T_{\mu\nu} \, ,
\end{equation}
which corresponds to the equation $M^2 I_R^{(1)} = M^2 I_T^{(1)}$ in (\ref{P1}). 
A solution of (\ref{GWM3}) is given by
\begin{equation}
\label{GWM4}
\zeta_{xx}=- \frac{1}{2}y^2 \, , \quad
\zeta_{yy}=- \frac{1}{2}x^2 \, , \quad
\zeta_{zz}= \frac{1}{2}y^2 \, , \quad
\zeta_{tt}=- \frac{1}{2}x^2 \, , \quad
\mbox{other components}=0 \, .
\end{equation}
We find that the connections read
\begin{align}
\label{GWM5}
& \Gamma^x_{xy}=\Gamma^x_{yx} = - \Gamma^y_{xx} = - \frac{1}{2}
y \kappa^2 B^2 \, , \quad
\Gamma^y_{xy}=\Gamma^y_{yx} = - \Gamma^x_{yy} = - \frac{1}{2}
x \kappa^2 B^2\, , \quad
\Gamma^z_{zy}=\Gamma^z_{yz} = - \Gamma^y_{zz} = \frac{1}{2}
y \kappa^2 B^2 \, , \nn
& \Gamma^t_{xt}=\Gamma^t_{tx} = \Gamma^x_{tt} = \frac{1}{2}
x \kappa^2 B^2 \, , \quad
\mbox{other components}=0 \, .
\end{align}
Because
\begin{equation}
\label{GWM6}
R^\lambda_{\ \mu\rho\nu}= - \Gamma^\lambda_{\mu\rho,\nu}
+ \Gamma^\lambda_{\mu\nu,\rho}
+ \mathcal{O} \left( \left( \kappa^2 B^2 \right)^2 \right) \, ,
\end{equation}
we obtain
\begin{align}
\label{GWM7}
& R^x_{\ yxy} = - R^x_{\ yyx}=-R^y_{\ xxy} = R^y_{\ xyx} =  \kappa^2 B^2\, ,
\quad
R^x_{\ txt} = - R^x_{\ ttx} = - R^t_{\ xtx} = R^t_{\ xxt} = \frac{1}{2}
\kappa^2 B^2 \, , \nn
& R^y_{\ zyz} = -R^y_{\ zzy} = R^z_{\ yzy} = - R^z_{\ yyz} = - \frac{1}{2}
\kappa^2 B^2 \, , \quad
\mbox{other components}=0 \, .
\end{align}
The above results are consistent with (\ref{curvatures}).
The expressions in (\ref{GWM4}) with (\ref{GWM2}) show that we
should require
\begin{equation}
\label{GWMa1}
\kappa^2 B^2 x^2 \, , \ \kappa^2 B^2 y^2 \ll 1 \, ,
\end{equation}
nor we need to consider the higher order terms with respect to
$\kappa^2 B^2$.
The gauge conditions in (\ref{H4}) can be explicitly written as,
\begin{align}
\label{H4B}
0=& \nabla^\mu h_{\mu x} = \partial^\mu h_{\mu x}
+ \frac{1}{2}\kappa^2 B^2 \left( y^2 \left( h_{xx} - h_{zx} \right)
+ x^2 \left( h_{yx} + h_{tx} \right)
   - y h_{xy} + x h_{yy} + x h_{tt} \right) \, , \nn
0=& \nabla^\mu h_{\mu y} = \partial^\mu h_{\mu y}
+ \frac{1}{2}\kappa^2 B^2 \left( y^2 \left( h_{xy} - h_{zy} \right)
+ x^2 \left( h_{yy} + h_{yx} \right)
+ y h_{xx} - x h_{yy} - y h_{zz}
\right) \, , \nn
0=& \nabla^\mu h_{\mu z} = \partial^\mu h_{\mu z}
+ \frac{1}{2}\kappa^2 B^2 \left( y^2 \left( h_{xz} - h_{zz} \right)
+ x^2 \left( h_{yz} + h_{tz} \right)
   - y h_{yz}
\right) \, , \nn
0=& \nabla^\mu h_{\mu t} = \partial^\mu h_{\mu t}
+ \frac{1}{2}\kappa^2 B^2 \left( y^2 \left( h_{xt} - h_{zt} \right)
+ x^2 \left( h_{yt} + h_{tt} \right)
   - x h_{tt}
\right) \, , \nn
0=& g^{\mu\nu} h_{\mu\nu} = h_{xx} + h_{yy} + h_{zz} - h_{tt}
+ \frac{1}{2}\kappa^2 B^2 y^2 \left( h_{xx} - h_{zz} \right)
+ \frac{1}{2}\kappa^2 B^2 y^2 \left( h_{yy} + h_{tt} \right)
\, .
\end{align}
The above equations indicate that there appear the longitudinal modes in
general.

\subsection{Propagation of Gravitational Wave and Scattering}

Because
\begin{align}
\label{GWM8}
\nabla^2 h_{\mu\nu} =& g^{\rho\sigma} \nabla_{\rho} \nabla_{\sigma} h_{\mu\nu}
= g^{\rho\sigma} \left( \partial_{\rho}  \nabla_{\sigma} h_{\mu\nu}
   - \Gamma^\tau_{\rho\sigma} \nabla_\tau h_{\mu\nu}
   - \Gamma^\tau_{\rho\mu} \nabla_\sigma h_{\tau\nu}
   - \Gamma^\tau_{\rho\nu} \nabla_\sigma h_{\mu\tau} \right) \nn
=& \left( \eta^{\rho\sigma} - \kappa^2 B^2 \zeta^{\rho\sigma} \right)
\partial_{\rho}  \partial_{\sigma} h_{\mu\nu}
+ \eta^{\rho\sigma} \left( \partial_\rho
\left( - \Gamma^\tau_{\sigma\mu} h_{\tau\nu} - \Gamma^\tau_{\sigma\nu}
h_{\mu\tau} \right)
   - \Gamma^\tau_{\rho\sigma} \partial_\tau h_{\mu\nu}
   - \Gamma^\tau_{\rho\mu} \partial_\sigma h_{\tau\nu}
   - \Gamma^\tau_{\rho\nu} \partial_\sigma h_{\mu\tau} \right) \nn
& + \mathcal{O} \left( \left(\kappa^2 B^2 \right)^2 \right) \, ,
\end{align}
we find the following explicit expressions, 
which corresponds to the equation $M^2 \kappa^2 I_R^{(2)}=M^2 \kappa^2 I_T^{(1,1)}$ 
in Eq.~(\ref{P2}), 
\begin{align}
\label{GWM9A}
\nabla^2 h_{xx} =& \eta^{\rho\sigma} \partial_\rho \partial_\sigma h_{xx}
   - \frac{\kappa^2 B^2}{2} \left( y^2 \left( - \partial_x^2 + \partial_z^2
\right)
   - x^2 \left( \partial_y^2 + \partial_t^2 \right) \right) h_{xx} \nn
& + \kappa^2 B^2 \left\{
h_{xx}  + 2 y \partial_y h_{xx}
+ \left( - y \partial_x + x \partial_y \right) \left( h_{xy} + h_{yx} \right)
+ x \partial_t \left( h_{xt} + h_{tx} \right) \right\}\, , \\
\label{GWM9B}
\nabla^2 h_{yy} =& \eta^{\rho\sigma} \partial_\rho \partial_\sigma h_{yy}
   - \frac{\kappa^2 B^2}{2} \left( y^2 \left( - \partial_x^2 + \partial_z^2
\right)
   - x^2 \left( \partial_y^2 + \partial_t^2 \right) \right) h_{yy} \nn
& + \kappa^2 B^2 \left\{
h_{yy} + 2 x \partial_x h_{yy}
+ \left( - x \partial_y + y \partial_x \right) \left( h_{xy} + h_{yx} \right)
   - y \partial_z \left( h_{yz} + h_{zy} \right) \right\}\, , \\
\label{GWM9C}
\nabla^2 h_{zz} =& \eta^{\rho\sigma} \partial_\rho \partial_\sigma h_{zz}
   - \frac{\kappa^2 B^2}{2} \left( y^2 \left( - \partial_x^2 + \partial_z^2
\right)
   - x^2 \left( \partial_y^2 + \partial_t^2 \right) \right) h_{zz}
+ \kappa^2 B^2 \left\{
   - h_{zz} - 2 y \partial_y h_{zz}
+ y \partial_z \left( h_{zy} + h_{yz} \right) \right\}\, , \\
\label{GWM9D}
\nabla^2 h_{xy} =& \eta^{\rho\sigma} \partial_\rho \partial_\sigma h_{xy}
   - \frac{\kappa^2 B^2}{2} \left( y^2 \left( - \partial_x^2 + \partial_z^2
\right)
   - x^2 \left( \partial_y^2 + \partial_t^2 \right) \right) h_{xy} \nn
& + \kappa^2 B^2 \left\{
h_{xy} + \left( x \partial_x + y \partial_y \right) h_{xy}
+ \left( - x \partial_y  + y \partial_x \right) \left( h_{xx} - h_{yy} \right)
+ x \partial_t h_{ty}
   - y \partial_z h_{xz} \right\}\, , \\
\label{GWM9E}
\nabla^2 h_{xz} =& \eta^{\rho\sigma} \partial_\rho \partial_\sigma h_{xz}
   - \frac{\kappa^2 B^2}{2} \left( y^2 \left( - \partial_x^2 + \partial_z^2
\right)
   - x^2 \left( \partial_y^2 + \partial_t^2 \right) \right) h_{xz} \nn
& + \kappa^2 B^2 \left\{
\left( - y \partial_x  + x \partial_y \right) h_{yz}
+ x \partial_t h_{tz} + y \partial_z h_{xy} \right\}\, , \\
\label{GWM9F}
\nabla^2 h_{yz} =& \eta^{\rho\sigma} \partial_\rho \partial_\sigma h_{yz}
   - \frac{\kappa^2 B^2}{2} \left( y^2 \left( - \partial_x^2 + \partial_z^2
\right)
   - x^2 \left( \partial_y^2 + \partial_t^2 \right) \right) h_{yz} \nn
& + \kappa^2 B^2 \left\{
\left( x \partial_x - y \partial_ y \right)h_{yz}
+ \left( - x \partial_y + y \partial_x  \right) h_{xz}
+ y \partial_z \left( h_{yy} - h_{zz} \right) \right\}\, .
\end{align}
On the other hand,
\begin{align}
\label{GWM10}
& \delta T_{\mu\nu} \equiv \frac{\partial T_{\mu\nu} }{\partial g_{\rho\sigma}}
h_{\rho\sigma}
\, , \quad
\delta T_{xx} = - \frac{1}{2} B^2 h_{yy} \, , \quad
\delta T_{yy} = - \frac{1}{2} B^2 h_{xx} \, , \quad
\delta T_{zz} = B^2 \left( - \frac{1}{2} h_{zz} + \frac{1}{2} \left( h_{xx}
+ h_{yy} \right) \right) \, , \nn
& \delta T_{xy} = \frac{1}{2} B^2  h_{xy} \, , \quad
\delta T_{xz} =  - \frac{1}{2} B^2 h_{xz} \, , \quad
\delta T_{yz} =  - \frac{1}{2} B^2 h_{yz} \, .
\end{align}
By combining (\ref{GWM9A}), (\ref{GWM9B}), (\ref{GWM9C}), (\ref{GWM9D}),
(\ref{GWM9E}), (\ref{GWM9F}), and (\ref{GWM10}), we find Eq.~(\ref{H6}) gives,
\begin{align}
\label{GWM11A}
0 =& \eta^{\rho\sigma} \partial_\rho \partial_\sigma h_{xx}
   - \frac{\kappa^2 B^2}{2} \left( y^2 \left( - \partial_x^2 + \partial_z^2
\right)
   - x^2 \left( \partial_y^2 + \partial_t^2 \right) \right) h_{xx} \nn
& + \kappa^2 B^2 \left\{
2 y \partial_y h_{xx}
+ \left( - y \partial_x + x \partial_y \right) \left( h_{xy} + h_{yx} \right)
+ x \partial_t \left( h_{xt} + h_{tx} \right) \right\} \nn
& + \frac{1}{2} \kappa^2 B^2 \left( - h_{xx} + h_{yy} + h_{tt} + h_{zz}
\right) \, , \\
\label{GWM11B}
0 =& \eta^{\rho\sigma} \partial_\rho \partial_\sigma h_{yy}
   - \frac{\kappa^2 B^2}{2} \left( y^2 \left( - \partial_x^2 + \partial_z^2
\right)
   - x^2 \left( \partial_y^2 + \partial_t^2 \right) \right) h_{yy} \nn
& + \kappa^2 B^2 \left\{
2 x \partial_x h_{yy}
+ \left( - x \partial_y + y \partial_x \right) \left( h_{xy} + h_{yx} \right)
   - y \partial_z \left( h_{yz} + h_{zy} \right) \right\} \nn
& + \frac{1}{2} \kappa^2 B^2 \left( h_{xx} - h_{yy} - h_{zz}
   - h_{tt} \right) \, , \\
\label{GWM11C}
0 =& \eta^{\rho\sigma} \partial_\rho \partial_\sigma h_{zz}
   - \frac{\kappa^2 B^2}{2} \left( y^2 \left( - \partial_x^2 + \partial_z^2
\right)
   - x^2 \left( \partial_y^2 + \partial_t^2 \right) \right) h_{zz}
+ \kappa^2 B^2 \left\{
   - 2 y \partial_y h_{zz}
+ y \partial_z \left( h_{zy} + h_{yz} \right) \right\} \nn
& + \frac{1}{2} \kappa^2 B^2 \left( h_{xx} - h_{yy} - h_{zz}
   - h_{tt} \right) \, , \\
\label{GWM11D}
0 =& \eta^{\rho\sigma} \partial_\rho \partial_\sigma h_{xy}
   - \frac{\kappa^2 B^2}{2} \left( y^2 \left( - \partial_x^2 + \partial_z^2
\right)
   - x^2 \left( \partial_y^2 + \partial_t^2 \right) \right) h_{xy} \nn
& + \kappa^2 B^2 \left\{
\left( x \partial_x + y \partial_y \right) h_{xy}
+ \left( - x \partial_y  + y \partial_x \right) \left( h_{xx} - h_{yy} \right)
+ x \partial_t h_{ty} - y \partial_z h_{xz}
\right\} - \kappa^2 B^2 h_{xy} \, , \\
\label{GWM11E}
0 =& \eta^{\rho\sigma} \partial_\rho \partial_\sigma h_{xz}
   - \frac{\kappa^2 B^2}{2} \left( y^2 \left( - \partial_x^2 + \partial_z^2
\right)
   - x^2 \left( \partial_y^2 + \partial_t^2 \right) \right) h_{xz} \nn
& + \kappa^2 B^2 \left\{ \left( - y \partial_x  + x \partial_y \right) h_{yz}
+ x \partial_t h_{tz} + y \partial_z h_{xy} \right\}
   - \kappa^2  B^2 h_{xz}\, , \\
\label{GWM11F}
0 =& \eta^{\rho\sigma} \partial_\rho \partial_\sigma h_{yz}
   - \frac{\kappa^2 B^2}{2} \left( y^2 \left( - \partial_x^2 + \partial_z^2
\right)
   - x^2 \left( \partial_y^2 + \partial_t^2 \right) \right) h_{yz} \nn
& + \kappa^2 B^2 \left\{
\left( x \partial_x - y \partial_ y \right)h_{yz}
+ \left( - x \partial_y + y \partial_x  \right) h_{xz}
+ y \partial_z \left( h_{yy} - h_{zz} \right) \right\} \, .
\end{align}

We investigate the propagation of the gravitational wave based on
the above equations.
In order to see the effect of the magnetic field, we assume
\begin{equation}
\label{GWM12}
h_{xy} = h^{(0)}_\times \sin k \left(z - t \right)
+ \mathcal{O}\left( \kappa^2 B^2 \right) \, , \quad
\mbox{other components} = \mathcal{O}\left( \kappa^2 B^2 \right) \, ,
\end{equation}
which corresponds to $\times$ mode propagating in parallel with the magnetic
field.
Then we obtain,
\begin{align}
\label{GWM13ABCF}
0 =& \eta^{\rho\sigma} \partial_\rho \partial_\sigma h_{xx}
= \eta^{\rho\sigma} \partial_\rho \partial_\sigma h_{yy}
= \eta^{\rho\sigma} \partial_\rho \partial_\sigma h_{zz}
= \eta^{\rho\sigma} \partial_\rho \partial_\sigma h_{yz}\, , \\
\label{GWM13D}
0 =& \eta^{\rho\sigma} \partial_\rho \partial_\sigma h_{xy}
+ \frac{\kappa^2 B^2 k^2 }{2} \left( y^2 - x^2 \right)
h^{(0)}_\times \sin k \left(z - t \right)
   - \kappa^2 B^2 h^{(0)}_\times \sin k \left(z - t \right) \, , \\
\label{GWM13E}
0 =& \eta^{\rho\sigma} \partial_\rho \partial_\sigma h_{xz}
+ \kappa^2 B^2 y k h^{(0)}_\times \cos k \left(z - t \right) \, , \\
\end{align}
Therefore if we define $\Box \equiv \eta^{\rho\sigma} \partial_\rho
\partial_\sigma$, we find
\begin{align}
\label{GWM14}
& h_{xx}=h_{yy}=h_{zz}=h_{yz}=0 \, , \quad
h_{xz} = - \kappa^2 B^2 k h^{(0)}_\times \Box^{-1}
\left( y\cos k \left(z - t \right) \right) \, , \nn
& h_{xy} = h^{(0)}_\times \left\{ \sin k \left(z - t \right)
   - \kappa^2 B^2 \Box^{-1} \left( \left( \frac{k^2 (y^2 - x^2)}2 - 1 \right)
\sin k \left(z - t \right) \right)
\right\} \, .
\end{align}
The $\mathcal{O}\left( \kappa^2 B^2 \right)$ is given by the scattering
of the gravitational wave by the magnetic field.
It could be interesting that there appears non-trivial $h_{xz}$ component.

Next we explore the $+$ mode propagating in parallel with
the magnetic field,
\begin{equation}
\label{GWM15}
h_{xx} = - h_{yy} = h^{(0)}_+ \sin k \left(z - t \right)
+ \mathcal{O}\left( \kappa^2 B^2 \right) \, , \quad
\mbox{other components} = \mathcal{O}\left( \kappa^2 B^2 \right) \, .
\end{equation}
Then we find
\begin{align}
\label{GWM16A}
0 =& \eta^{\rho\sigma} \partial_\rho \partial_\sigma h_{xx}
+ \frac{\kappa^2 B^2 k^2}{2} \left( y^2 - x^2 \right)
h^{(0)}_+ \sin k \left(z - t \right)
+ \kappa^2 B^2 h^{(0)}_+ \sin k \left(z - t \right) \, , \\
\label{GWM1B}
0 =& - \eta^{\rho\sigma} \partial_\rho \partial_\sigma h_{yy}
+ \frac{\kappa^2 B^2 k^2}{2} \left( y^2 - x^2 \right)
h^{(0)}_+ \sin k \left(z - t \right)
+ \kappa^2 B^2 h^{(0)}_+ \sin k \left(z - t \right) \, , \\
\label{GWM16CDE}
0 =& \eta^{\rho\sigma} \partial_\rho \partial_\sigma h_{zz}
= \eta^{\rho\sigma} \partial_\rho \partial_\sigma h_{xy}
= \eta^{\rho\sigma} \partial_\rho \partial_\sigma h_{xz} \, , \\
\label{GWM16F}
0 =& \eta^{\rho\sigma} \partial_\rho \partial_\sigma h_{yz}
   - \kappa^2 B^2 y k h^{(0)}_+ \cos k \left(z - t \right) \, ,
\end{align}
and the solution is given by
\begin{align}
\label{GWM17}
& h_{zz}=h_{xy}=h_{xz}=0 \, , \quad
h_{yz} = + \kappa^2 B^2 k h^{(0)}_+ \Box^{-1}
\left( y\cos k \left(z - t \right) \right) \, , \nn
& h_{xx} = - h_{yy} = h^{(0)}_+ \left\{ \sin k \left(z - t \right)
   - \kappa^2 B^2 \Box^{-1} \left( \left( \frac{k^2 (y^2 - x^2)}2 + 1 \right)
\sin k \left(z - t \right) \right)
\right\} \, .
\end{align}
Then there appears non-trivial $h_{yz}$ component.
The physical behavior of the $+$ mode (\ref{GWM17}) does not change
from that of the $\times$ mode in (\ref{GWM14}).

We examine the $\times$ mode propagating perpendicular to
the magnetic field,
\begin{equation}
\label{GWM18}
h_{xz} = {\bar h}^{(0)}_\times \sin k \left(y - t \right)
+ \mathcal{O}\left( \kappa^2 B^2 \right) \, , \quad
\mbox{other components} = \mathcal{O}\left( \kappa^2 B^2 \right) \, .
\end{equation}
Then we find
\begin{align}
\label{GWM19ABCD}
0 =& \eta^{\rho\sigma} \partial_\rho \partial_\sigma h_{xx}
= \eta^{\rho\sigma} \partial_\rho \partial_\sigma h_{yy}
= \eta^{\rho\sigma} \partial_\rho \partial_\sigma h_{zz}
= \eta^{\rho\sigma} \partial_\rho \partial_\sigma h_{xy} \, , \\
\label{GWM19E}
0 =& \eta^{\rho\sigma} \partial_\rho \partial_\sigma h_{xz}
   - \kappa^2 B^2 k^2 x^2 {\bar h}^{(0)}_\times \sin k \left(y - t \right)
   - \kappa^2  B^2 {\bar h}^{(0)}_+ \sin k \left(y - t \right) \, , \\
\label{GWM19F}
0 =& \eta^{\rho\sigma} \partial_\rho \partial_\sigma h_{yz}
   - \kappa^2 B^2 k {\bar h}^{(0)}_\times \cos k \left(y - t \right)\, ,
\end{align}
whose solution is given by
\begin{align}
\label{GWM20}
& h_{xx} = h_{yy} = h_{zz} = h_{xy} = 0 \, , \quad
h_{xz} = {\bar h}^{(0)}_\times \left\{ \sin k \left(y - t \right)
+ \kappa^2 B^2 \Box^{-1} \left( \left( k^2 x^2 + 1 \right)
\sin k \left(y - t \right) \right) \right\}\, , \nn
& h_{yz} = \kappa^2 B^2 k {\bar h}^{(0)}_\times
\Box^{-1} \left( x \cos k \left(y - t \right) \right)\, .
\end{align}

In case of the $+$ mode propagating perpendicular to
the magnetic field,
\begin{equation}
\label{GWM21}
h_{xx} = - h_{zz} = {\bar h}^{(0)}_+ \sin k \left(y - t \right)
+ \mathcal{O}\left( \kappa^2 B^2 \right) \, , \quad
\mbox{other components} = \mathcal{O}\left( \kappa^2 B^2 \right) \, ,
\end{equation}
we obtain
\begin{align}
\label{GWM22A}
0 =& \eta^{\rho\sigma} \partial_\rho \partial_\sigma h_{xx}
   - \kappa^2 B^2 k^2 x^2 {\bar h}^{(0)}_+ \sin k \left(y - t \right)
+ 2 \kappa^2 B^2 k y {\bar h}^{(0)}_+ \cos k \left(y - t \right)
   -  \kappa^2 B^2 {\bar h}^{(0)}_+ \sin k \left(y - t \right) \, , \\
\label{GWM22BEF}
0 =& \eta^{\rho\sigma} \partial_\rho \partial_\sigma h_{yy}
= \eta^{\rho\sigma} \partial_\rho \partial_\sigma h_{xz}
=  \eta^{\rho\sigma} \partial_\rho \partial_\sigma h_{yz} \, , \\
\label{GWM22C}
0 =& \eta^{\rho\sigma} \partial_\rho \partial_\sigma h_{zz}
+ \kappa^2 B^2 k^2 x^2  {\bar h}^{(0)}_+ \sin k \left(y - t \right)
+ 2 \kappa^2 B^2 k y {\bar h}^{(0)}_+ \cos k \left(y - t \right)
+ \kappa^2 B^2 {\bar h}^{(0)}_+ \sin k \left(y - t \right) \, , \\
\label{GWM22D}
0 =& \eta^{\rho\sigma} \partial_\rho \partial_\sigma h_{xy}
   - \kappa^2 B^2 k x {\bar h}^{(0)}_+ \cos k \left(y - t \right) \, .
\end{align}
Then we find
\begin{align}
\label{GWM23}
h_{xx} =& {\bar h}^{(0)}_+ \left\{ \sin k \left(y - t \right)
+ \kappa^2 B^2 \Box^{-1} \left( \left( k^2 x^2 + 1 \right)
\sin k \left(y - t \right)
   - 2 k y \cos k \left(y - t \right) \right) \right\} \, , \quad
h_{yy} = h_{xz} = h_{yz} =0 \, , \nn
h_{zz} =& {\bar h}^{(0)}_+ \left\{ - \sin k \left(y - t \right)
+ \kappa^2 B^2 \Box^{-1} \left( - \left( k^2 x^2 + 1 \right)
\sin k \left(y - t \right)
   - 2 k y \cos k \left(y - t \right) \right) \right\} \, , \nn
h_{xy} =& \kappa^2 B^2 k {\bar h}^{(0)}_+ \Box^{-1} \left(
x \cos k \left(y - t \right) \right) \, .
\end{align}
The behavior of the $+$ mode in (\ref{GWM23})
seems to be rather different from
that of the $\times$ mode (\ref{GWM20}).

In the above expressions, in order that the perturbation should be
consistent, in addition to (\ref{GWMa1}), we need to require
\begin{equation}
\label{GWMa2}
\frac{\kappa^2 B^2}{k^2} \ll 1 \, .
\end{equation}

We should note that $\Box^{-1}$ is the retarded propagator, which is
non-local and satisfies the equation
\begin{equation}
\label{GWM24}
\Box G\left( x^\mu, {x'}^\mu \right) = \delta^4 \left( x^\mu - {x'}^\mu
\right) \, , \quad G\left( x^\mu, {x'}^\mu \right) \equiv \Box^{-1} \, .
\end{equation}
Therefore for any function $f\left( x^\mu \right)$ of the space-time
coordinate $x^\mu$, we have
\begin{equation}
\label{GWM25}
\Box^{-1} f\left( x^\mu \right)
\equiv \int_V d^4 x' G\left( x^\mu, {x'}^\mu \right)
f\left( {x'}^\mu \right) \, .
\end{equation}
The region $V$ of the integration is given by the region of the space-time,
where the magnetic field exists.
Therefore the gravitational wave carries the information of the distribution
of the magnetic field in the universe.
In the above analysis, we have assumed that $\kappa^2 B^2$ should be
small enough.
Even if $\kappa^2 B^2$ is small, the integration over the space-time in
(\ref{GWM25}) enhance the amplitude of the gravitational wave.

As an example, we consider NGC~6946, which is a spiral galaxy.
The size of NGC~6946 is $\sim 100\, \mbox{k light years} \sim 10^{28}/
\mathrm{eV}$ and its distance from the earth is $20\, \mbox{M light year}
\sim 10^{30}/\mathrm{eV}$.
We may estimate,
\begin{equation}
\label{GWM26}
\int_V d^4 x' G\left( x^\mu, {x'}^\mu \right)
f\left( {x'}^\mu \right) \sim
\left( 10^{30}/\mathrm{eV} \right)^{-1}
\left( 10^{28}/\mathrm{eV} \right)^3 f
= 10^{54}/\mathrm{eV}^2f  \, .
\end{equation}
The exponents $-1$ and $3$ come because we are considering the static
magnetic field.
Because NGC~6946 has a magnetic field with $\mu$G, we
may evaluate $B^2$ as
\begin{equation}
\label{GWM27}
B^2 \sim 10^{-16}\, \mathrm{eV}^4 \, ,
\end{equation}
and therefore
\begin{equation}
\label{GWM28}
\kappa^2 B^2 \sim 10^{-72}\, \mathrm{eV}^2 \, .
\end{equation}
We may also estimate $x$ and $y$ from the size of the galaxy as
\begin{equation}
\label{GWM28B}
x \sim y \sim 10^{28}/\mathrm{eV} \, .
\end{equation}
Therefore we obtain,
\begin{equation}
\label{GWMa3}
\kappa^2 B^2 x^2 \sim \kappa^2 B^2 y^2 \sim 10^{-16} \ll 1 \, ,
\end{equation}
and as a result the condition (\ref{GWMa1}) is satisfied.

In case of the gravitational wave GW150914, the typical frequency is
100-500 Hz, which corresponds to the wave number
\begin{equation}
\label{GWM29}
k \sim 10^{-6} \, \mathrm{m}^{-1} \sim 10^{-13}\, \mathrm{eV}\, ,
\end{equation}
and hence
\begin{equation}
\label{GWM30}
kx \sim ky \sim 10^{22} \, .
\end{equation}
Because
\begin{equation}
\label{GWMa4}
\frac{\kappa^2 B^2}{k^2} \sim 10^{-46} \ll 1 \, ,
\end{equation}
the condition (\ref{GWMa2}) is also satisfied.

For example, in (\ref{GWM14}), we find,
\begin{align}
\label{GWM31}
h_{xz} =& - \kappa^2 B^2 k h^{(0)}_\times \Box^{-1}
\left( y\cos k \left(z - t \right) \right)
\sim 10^{-72+54+22} h^{(0)}_\times \cos k \left(z - t \right)
= 10^4 h^{(0)}_\times \cos k \left(z - t \right) \, , \nn
h_{xy} =& h^{(0)}_\times \left\{ \sin k \left(z - t \right)
   - \kappa^2 B^2 \Box^{-1} \left( \left( \frac{k^2 (y^2 - x^2)}2 - 1 \right)
\sin k \left(z - t \right) \right) \right\}
\sim \left( 1 + 10^{26} \right)
h^{(0)}_\times \sin k \left(z - t \right) \, ,
\end{align}
which shows that the corrections are rather large.
The large correction comes from the large size of the galaxy.
Of course, we have neglected the numerical factors and we
have assumed that the magnetic field is uniform at the large
scale of the galaxy, which indicates that we should have overestimated
the value.
Furthermore the correction to the wave number is enhanced because it is
multiplied by the distance of the propagation, which gives a
non-negligible
correction in the phase. However, we expand the expression with respect
to $\kappa^2 B^2$, there appears the large correction,
for example, $\sin \left( \left( k + \alpha \kappa^2 B^2 \right) z \right)
\sim \sin \left( k z \right) + \cos \left( kz \right) \alpha \kappa^2 B^2 z$,
where we express the correction by $\alpha \kappa^2 B^2 $, which might
be small but $\alpha \kappa^2 B^2 z$ (multiplied with $z$) is not small
in general.
Anyway the magnetic field may give a contribution which
cannot be neglected.

We may study magnetar~\cite{Duncan:1992hi}, which has very strong
magnetic field of $10^{11}\, \mathrm{T}$. 
Then we find $B^2\sim 10^{26}\, \mathrm{eV}$ and therefore
$\kappa^2 B^2 \sim 10^{-30}\, \mathrm{eV}^2$.
If we consider the
gravitational wave in (\ref{GWM29}), we find
\begin{equation}
\label{GWMa4B}
\frac{\kappa^2 B^2}{k^2} \sim 10^{-4} \ll 1 \, ,
\end{equation}
which is still small and the condition (\ref{GWMa2}) is also satisfied.
Thus the perturbation is still valid.
The typical size of the magnetic field in magnetar is
$2\times 10^8\, \mathrm{km} \sim 10^{15}/\mathrm{eV}$ and therefore
\begin{equation}
\label{magskxky}
kx \sim ky \sim 10^3 \, ,
\end{equation}
and
\begin{equation}
\label{GWMa3B}
\kappa^2 B^2 x^2 \sim \kappa^2 B^2 y^2 \sim 1 \, ,
\end{equation}
which tells that the condition (\ref{GWMa1}) is not always satisfied.
In case of SGR 1806-20, which is the first  magnetar found, the distance
from
the earth is $5\times 10^{4}$ light years $\sim$ $10^{27}/\mathrm{eV}$,
instead of (\ref{GWM26}), we may estimate,
\begin{equation}
\label{GWM26B}
\int_V d^4 x' G\left( x^\mu, {x'}^\mu \right)
f\left( {x'}^\mu \right) \sim
\left( 10^{27}/\mathrm{eV} \right)^{-1}
\left( 10^{15}/\mathrm{eV} \right)^3 f
= 10^{18}/\mathrm{eV}^2f  \, .
\end{equation}
The expressions corresponding to (\ref{GWM31}) have the following form,
\begin{align}
\label{GWM31B}
h_{xz} =& - \kappa^2 B^2 k h^{(0)}_\times \Box^{-1}
\left( y\cos k \left(z - t \right) \right)
\sim 10^{-30+3+18} h^{(0)}_\times \cos k \left(z - t \right)
= 10^{-9} h^{(0)}_\times \cos k \left(z - t \right) \, , \nn
h_{xy} =& h^{(0)}_\times \left\{ \sin k \left(z - t \right)
   - \kappa^2 B^2 \Box^{-1} \left( \left( \frac{k^2 (y^2 - x^2)}2 - 1 \right)
\sin k \left(z - t \right) \right) \right\}
\sim \left( 1 + 10^{-6} \right)
h^{(0)}_\times \sin k \left(z - t \right) \, ,
\end{align}
where the corrections are reasonably small.Hence, contrary to the case of
the
large magnetic field, it could be difficult to detect the effect of the
magnetic field.

\subsection{Propagation of Gravitational Wave by Adiabatic Approximation}

In the last subsection, we have investigated the scattering
of the gravitational wave by the magnetic field.
As mentioned after Eq.~(\ref{em2}), there are two kinds of sources for the
scattering.
One is given by the change of the geometry and another is the
fluctuation of the distribution of the magnetic field given by the
gravitational wave.
The obtained results seem to say that the effects in the change of the
geometry are much larger than those of the fluctuation of the magnetic field.

The expressions (\ref{GWM14}), (\ref{GWM17}), and (\ref{GWM23})
(except of (\ref{GWM21})) could not be valid for the magnetic field of
the
galactic size although they may be valid for the magnetic field with smaller
size.
This could be mainly because the variation of the phase is not small
although the correction of the wave number is small.
Then we now try to solve Eqs.~(\ref{GWM11A}),
(\ref{GWM11B}), (\ref{GWM11C}), (\ref{GWM11D}), (\ref{GWM11E}),
and (\ref{GWM11F}) by using the adiabatic approximation, where we
neglect the derivative with respect to the background.

First we analyze the gravitational wave along $z$-axis and assume
\begin{equation}
\label{GWM33}
h_{ij} = h_{ij}^{(0)} \left( x,y \right) \e^{i\left( k \left( x,y \right) z
   - \omega \left( x,u \right) t \right)}\, , \quad h_{ti}=h_{it}=h_{tt}=0 \, .
\end{equation}
Finally we choose the real part of the above expression.
We take $h_{ij}^{(0)} \left( x,y \right)$, $k \left( x,y \right)$, and
$\omega \left( x,u \right)$ can depend on the coordinates $(x,y)$
although we neglect the derivative with respect to $x$ and $y$ for
$h_{ij}^{(0)} \left( x,y \right)$, $k \left( x,y \right)$, and
$\omega \left( x,u \right)$.
Then Eqs.~(\ref{GWM11A}),
(\ref{GWM11B}), (\ref{GWM11C}), (\ref{GWM11D}), (\ref{GWM11E}),
and (\ref{GWM11F}) give the following algebraic equation:
\begin{align}
\label{GWM34A}
0 =& \left( - k^2 + \omega^2 - \frac{\kappa^2 B^2}{2}
\left( - k^2 y^2 + \omega^2 x^2 \right) \right) h_{xx}^{(0)}
+ \frac{1}{2} \kappa^2 B^2 \left( - h_{xx}^{(0)} + h_{yy}^{(0)}
+ h_{zz}^{(0)} \right) \, , \\
\label{GWM34B}
0 =& \left( - k^2 + \omega^2 - \frac{\kappa^2 B^2}{2}
\left( -k^2 y^2 + \omega^2 x^2 \right) \right) h_{yy}^{(0)}
- 2 i k y \kappa^2 B^2 h_{yz}^{(0)}
+ \frac{1}{2} \kappa^2 B^2 \left( h_{xx}^{(0)} - h_{yy}^{(0)}
   - h_{zz}^{(0)} \right) \, , \\
\label{GWM34C}
0 =& \left( - k^2 + \omega^2 - \frac{\kappa^2 B^2}{2}
\left( - k^2 y^2 + \omega^2 x^2 \right) \right) h_{zz}^{(0)}
+ 2i k y h_{yz}^{(0)} + \frac{1}{2} \kappa^2 B^2
\left( h_{xx}^{(0)} - h_{yy}^{(0)} - h_{zz}^{(0)} \right) \, , \\
\label{GWM34D}
0 =& \left( - k^2 + \omega^2 - \frac{\kappa^2 B^2}{2}
\left( - k^2 y^2 + \omega^2 x^2 \right) \right) h_{xy}^{(0)}
   - ik y \kappa^2 B^2 h_{xz}^{(0)} - \kappa^2 B^2 h_{xy}^{(0)} \, , \\
\label{GWM34E}
0 =& \left( - k^2 + \omega^2 - \frac{\kappa^2 B^2}{2}
\left( - k^2 y^2 + \omega^2 x^2 \right) \right) h_{xz}^{(0)}
+ iky \kappa^2 B^2 h_{xy}^{(0)} - \kappa^2  B^2 h_{xz}^{(0)}\, , \\
\label{GWM34F}
0 =& \left( - k^2 + \omega^2 - \frac{\kappa^2 B^2}{2}
\left( - k^2 y^2 + \omega^2 x^2 \right) \right) h_{yz}^{(0)}
+ iky \kappa^2 B^2 \left( h_{yy}^{(0)} - h_{zz}^{(0)} \right) \, .
\end{align}

Then the solution corresponding to the $+$ mode is given by
\begin{equation}
\label{GWM35}
h_{xx}^{(0)} = - h_{yy}^{(0)}\, , \quad \mbox{other components}=0 \, ,
\end{equation}
with the dispersion relation,
\begin{equation}
\label{GWM36}
0 = - k^2 + \omega^2 - \frac{\kappa^2 B^2}{2}
\left( - k^2 y^2 + \omega^2 x^2 \right) - \kappa^2 B^2 \, .
\end{equation}
As clear from the dispersion relation, there appears the square of
an effective mass
$\sim \kappa^2 B^2$ in addition to the $x,y$ dependent shift of the
phase.
Eq.~(\ref{GWM36}) leads to
\begin{equation}
\label{GWM36B}
\left( 1 + \frac{\kappa^2B^2 y^2}{2} \right)
= \left(1 + \frac{\kappa^2 B^2 x^2}{2} \right) \omega^2 - \kappa^2 B^2 \, ,
\end{equation}
or
\begin{equation}
\label{GWM36C}
k^2 = \left( 1 + \frac{\kappa^2 B^2 \left( x^2 - y^2 \right) }{2} \right)
\omega^2 - \kappa^2 B^2 + \mathcal{O} \left( \left( \kappa^2 B^2 \right)^2
\right)
\, .
\end{equation}
The factor $1 + \frac{\kappa^2 B^2 \left( x^2 - y^2 \right)}{2}$ in front of
$\omega^2$ comes from the change of the geometry and therefore
there appears the same correction even for the propagation of light but
the term $\kappa^2 B^2$ gives the square
of the effective mass, which is absent in the photon.

The solution corresponding to the $\times$ mode is given by,
\begin{equation}
\label{GWM37}
h_{xz}^{(0)} =\frac{1}{iky\kappa^2 B^2}  \left( - k^2 + \omega^2 -
\frac{\kappa^2 B^2}{2}
\left( - k^2 y^2 + \omega^2 x^2 \right) - \kappa^2 B^2
\right) h_{xy}^{(0)} \, ,
\quad \mbox{other components}=0 \, ,
\end{equation}
with a little bit complex dispersion relation,
\begin{equation}
\label{GWM38}
\left( - k^2 + \omega^2 - \frac{\kappa^2 B^2}{2}
\left( - k^2 y^2 + \omega^2 x^2 \right) - \kappa^2 B^2 \right)^2
= k^2 y^2 \kappa^4 B^4 \, ,
\end{equation}
that is
\begin{equation}
\label{GWM39}
0 = - k^2 + \omega^2 - \frac{\kappa^2 B^2}{2}
\left( - k^2 y^2 + \omega^2 x^2 \right) - \kappa^2 B^2
\pm k y \kappa^2 B^2 \, .
\end{equation}
We should note that there should appear $h_{xz}$ component whose
phase is different from that of $h_{xy}$ component by $\frac{\pi}{2}$.

Next we consider the gravitational wave along $y$-axis and assume
\begin{equation}
\label{GWM40}
h_{ij} = h_{ij}^{(0)} \left( x,y \right) \e^{i\left( k \left( x,y \right) y
   - \omega \left( x,u \right) t \right)}\, , \quad h_{ti}=h_{it}=h_{tt}=0 \, .
\end{equation}
Again we neglect the derivative with respect to $x$ and $y$ for
$h_{ij}^{(0)} \left( x,y \right)$, $k \left( x,y \right)$, and
$\omega \left( x,u \right)$ as an adiabatic approximation.
Then we find the following equations,
\begin{align}
\label{GWM41A}
0 =& \left( - k^2 + \omega^2  - \frac{\kappa^2 B^2 x^2}{2}
\left( k^2 + \omega^2 \right) \right) h_{xx}^{(0)}
+ 2i k \kappa^2 B^2 \left( y h_{xx}^{(0)} + x h_{xy}^{(0)} \right)
+ \frac{1}{2} \kappa^2 B^2 \left( - h_{xx}^{(0)} + h_{yy}^{(0)} + h_{zz}^{(0)}
\right) \, , \\
\label{GWM41B}
0 =& \left( - k^2 + \omega^2  - \frac{\kappa^2 B^2 x^2}{2}
\left( k^2 + \omega^2 \right) \right) h_{yy}^{(0)}
   - 2 i k  \kappa^2 B^2 x h_{xy}^{(0)}
+ \frac{1}{2} \kappa^2 B^2 \left( h_{xx}^{(0)} - h_{yy}^{(0)} - h_{zz}^{(0)}
\right) \, , \\
\label{GWM41C}
0 =& \left( - k^2 + \omega^2  - \frac{\kappa^2 B^2 x^2}{2}
\left( k^2 + \omega^2 \right) \right) h_{zz}^{(0)}
   - 2i k \kappa^2 B^2 y h_{zz}^{(0)}
+ \frac{1}{2} \kappa^2 B^2 \left( h_{xx}^{(0)} - h_{yy}^{(0)} - h_{zz}^{(0)}
\right) \, , \\
\label{GWM41D}
0 =& \left( - k^2 + \omega^2  - \frac{\kappa^2 B^2 x^2}{2}
\left( k^2 + \omega^2 \right) \right) h_{xy}^{(0)}
+ ik \kappa^2 B^2 \left\{ y h_{xy}^{(0)}  - x \left( h_{xx}^{(0)} -
h_{yy}^{(0)} \right) \right\} - \kappa^2 B^2 h_{xy}^{(0)} \, , \\
\label{GWM41E}
0 =& \left( - k^2 + \omega^2  - \frac{\kappa^2 B^2 x^2}{2}
\left( k^2 + \omega^2 \right) \right) h_{xz}^{(0)}
+ i k \kappa^2 B^2 x h_{yz}^{(0)} - \kappa^2  B^2 h_{xz}^{(0)}\, , \\
\label{GWM41F}
0 =& \left( - k^2 + \omega^2  - \frac{\kappa^2 B^2 x^2}{2}
\left( k^2 + \omega^2 \right) \right) h_{yz}^{(0)}
+ ik \kappa^2 B^2 \left( - y h_{yz}^{(0)}
   - x h_{xz}^{(0)} \right) \, .
\end{align}
Not as in Eqs.~(\ref{GWM34A}),
(\ref{GWM34B}), (\ref{GWM34C}), (\ref{GWM34D}), (\ref{GWM34E}),
and (\ref{GWM34F}), it is difficult to solve Eqs.~(\ref{GWM41A}),
(\ref{GWM41B}), (\ref{GWM41C}), (\ref{GWM41D}), (\ref{GWM41E}),
and (\ref{GWM41F}).
First we assume $h_{xz}^{(0)} = h_{yz}^{(0)} = 0$ and the following
dispersion relation,
\begin{equation}
\label{GWM42}
0 = - k^2 + \omega^2  - \frac{\kappa^2 B^2 x^2}{2}
\left( k^2 + \omega^2 \right) + 2i k \kappa^2 B^2 y  - \kappa^2 B^2
+ \lambda \kappa^2 B^2 \, .
\end{equation}
By substituting (\ref{GWM42}) into Eqs.~(\ref{GWM41A}),
(\ref{GWM41B}), (\ref{GWM41C}), (\ref{GWM41D}), we obtain
\begin{align}
\label{GWM41Ab}
0 =& 2i k x h_{xy}^{(0)}
+ \frac{1}{2} \left( h_{xx}^{(0)} + h_{yy}^{(0)} + h_{zz}^{(0)}
\right) + \lambda h_{xx}^{(0)} \, , \\
\label{GWM41Bb}
0 =&  - 2 i k \left( y h_{yy}^{(0)} + x h_{xy}^{(0)} \right)
+ \frac{1}{2} \left( h_{xx}^{(0)} + h_{yy}^{(0)} - h_{zz}^{(0)}
\right) + \lambda h_{yy}^{(0)} \, , \\
\label{GWM41Cb}
0 =& - 4i k y h_{zz}^{(0)}
+ \frac{1}{2} \left( h_{xx}^{(0)} - h_{yy}^{(0)} + h_{zz}^{(0)} \right)
   + \lambda h_{zz}^{(0)}\, , \\
\label{GWM41Db}
0 =& ik \left\{ - y h_{xy}^{(0)}  - x \left( h_{xx}^{(0)} - h_{yy}^{(0)}
\right) \right\}
+ \lambda h_{xy}^{(0)} \, ,
\end{align}
which can be rewritten by using a matrix,
\begin{equation}
\label{GWM41c}
0 = \left( \begin{array}{cccc}
\lambda + \frac{1}{2} & \frac{1}{2} & \frac{1}{2} & 2ikx \\
\frac{1}{2} & \lambda + \frac{1}{2} - 2 i k y & - \frac{1}{2}
& - 2 i k x \\
\frac{1}{2} & - \frac{1}{2} & \frac{1}{2} - 4iky + \lambda & 0 \\
   - ik x & ikx & 0 & -iky + \lambda
\end{array} \right)
\left( \begin{array}{c}
h_{xx}^{(0)} \\ h_{yy}^{(0)} \\ h_{zz}^{(0)} \\ h_{xy}^{(0)}
\end{array} \right) \, .
\end{equation}
Then the dispersion relation (\ref{GWM42}) can be determined by solving the
following equation,
\begin{equation}
\label{GWM41d}
0 = \left| \begin{array}{cccc}
\lambda + \frac{1}{2} & \frac{1}{2} & \frac{1}{2} & 2ikx \\
\frac{1}{2} & \lambda + \frac{1}{2} - 2 i k y & - \frac{1}{2}
& - 2 i k x \\
\frac{1}{2} & - \frac{1}{2} & \frac{1}{2} - 4iky + \lambda & 0 \\
   - ik x & ikx & 0 & -iky + \lambda
\end{array} \right| \, .
\end{equation}

In case that $\left|kx\right|, \left|ky\right| \ll 1$, we may approximate
Eq.~(\ref{GWM41c}) as follows,
\begin{equation}
\label{GWM41dd}
0 = \lambda \left\{ \left( \lambda + \frac{1}{2} \right)^3
- \frac{3}{4} \left( \lambda + \frac{1}{2} \right)
- \frac{1}{4} \right\}
= \lambda \left( \lambda - \frac{1}{2} \right) \left( \lambda + 1 \right)^2
\, ,
\end{equation}
whose solution is given by, $\lambda=0$, $\frac{1}{2}$, and two $\lambda=-1$.
Because there appears the imaginary part in the dispersion relation
(\ref{GWM42}), the amplitude of the gravitational wave is decaying.
The mode corresponding to $\lambda=0$ gives
$h_{xx}^{(0)} = h_{yy}^{(0)} = h_{zz}^{(0)} =0$ and $h_{xy}^{(0)} \neq 0$
and therefore the mode could be unphysical.
The mode corresponding to $\lambda=\frac{1}{2}$ gives
$h_{yy}^{(0)} = h_{zz}^{(0)} = - h_{xx}^{(0)}$ and the mode to $\lambda=-1$
gives $h_{xx}^{(0)} = h_{yy}^{(0)} + h_{zz}^{(0)}$.
These are not connected with the modes of the gravitational wave in the vacuum.
In case that $\left|kx\right|, \left|ky\right| \gg 1$, by using
Eq.~(\ref{GWM41c}), we acquire
\begin{equation}
\label{GWM45}
0 =\left( \lambda - 4iky \right) \left( \lambda - iky \right)
\left( \lambda^2 - 2iky \lambda - 2k^2 x^2 \right)
\, ,
\end{equation}
whose solution is given by
\begin{equation}
\label{GWM46}
\lambda = iky\, , \ 4iky \, , \ iky \pm i k \sqrt{ y^2 - 2 x^2 } \, .
\end{equation}
For $\lambda = iky$, we obtain $h_{yy}^{(0)} = h_{yy}^{(0)}$,
$h_{zz}^{(0)} = 0$, and $h_{xy}^{(0)} = - \frac{y}{2x} h_{xx}^{(0)}$.
When $\lambda=4iky$, $h_{xx}^{(0)} = h_{yy}^{(0)} = h_{xy}^{(0)}=0$
and if $\lambda = iky \pm i k \sqrt{ y^2 - 2 x^2 }$, we find
$h_{zz}^{(0)} = 0$ and $h_{xy}^{(0)} = \frac{i\lambda}{2kx} h_{xx}^{(0)}$,
$h_{yy}^{(0)} = - \frac{\lambda}{\lambda - 2iky}  h_{xx}^{(0)}$.

We may also investigate the mode where
$h_{xx}^{(0)} = h_{yy}^{(0)} = h_{zz}^{(0)} = h_{xy}^{(0)} = 0$.
Then by using (\ref{GWM41E}) and (\ref{GWM41F}),
the dispersion relation is given by
\begin{equation}
\label{GWM47}
0 = \left( - k^2 + \omega^2  - \frac{\kappa^2 B^2 x^2}{2}
\left( k^2 + \omega^2 \right)  - \kappa^2 B^2 \right)
\left( - k^2 + \omega^2  - \frac{\kappa^2 B^2 x^2}{2}
\left( k^2 + \omega^2 \right) - ik y \kappa^2 B^2 \right)
   - k^2 \kappa^4 B^4 x^2 \, ,
\end{equation}
and we obtain
\begin{equation}
\label{GWM48}
h_{yz}^{(0)}  = - \frac{i}{k \kappa^2 B^2 x}
\left( - k^2 + \omega^2  - \frac{\kappa^2 B^2 x^2}{2}
\left( k^2 + \omega^2 \right) - \kappa^2 B^2 \right) h_{xz}^{(0)} \, .
\end{equation}
Then we find that the adiabatic approximation gives reasonable results for the large magnetic
field compared with simple perturbation with respect to $\kappa^2 B^2$ even if
$\kappa^2 B^2$ is small.
As mentioned after Eq.~(\ref{GWM31}), the overestimated amplitude in the
simple perturbation can be absorbed ito the phase and therefore the real amplitude
does not become so large.

\section{$F(R)$ gravity case \label{SecIV}}

Let us briefly discuss  $F(R)$ gravity in the similar context. Its
action is given by,
\begin{equation}
\label{JGRG7}
S_{F(R)}= \int d^4 x \sqrt{-g} \left(
\frac{F(R)}{2\kappa^2}
+ \mathcal{L}_\mathrm{matter} \left( g_{\mu\nu}, \Psi_i \right)
\right)\, ,
\end{equation}
where $\Psi_i$ expresses the field corresponding to matters.
It is well-known that by using the scale transformation,
\begin{equation}
\label{JGRG22}
g_{\mu\nu}\to \e^\sigma g_{\mu\nu}\, ,\quad \sigma = -\ln F'(A)\, ,
\end{equation}
the action (\ref{JGRG7}) can be rewritten in the
scalar-tensor form:
\begin{align}
\label{JGRG23}
S_E =& \int d^4 x \sqrt{-g} \left\{\frac{1}{2\kappa^2}\left(
R - \frac{3}{2}g^{\rho\sigma}
\partial_\rho \sigma \partial_\sigma \sigma - V(\sigma) \right)
+ \mathcal{L}_\mathrm{matter}
\left( \e^\sigma g_{\mu\nu}, \Psi_i \right)
\right\} \, ,\nn
V(\sigma) =& \e^\sigma g\left(\e^{-\sigma}\right)
   - \e^{2\sigma} f\left(g\left(\e^{-\sigma}\right)\right)\, .
\end{align}
Here $g\left(\e^{-\sigma}\right)$ is given by solving the equation
$\sigma = -\ln\left( 1 + f'(A)\right)=- \ln F'(A)$ as
$A=g\left(\e^{-\sigma}\right)$.
In case of the Einstein gravity in this paper, we have neglected the
cosmic expansion.
Hence we may assume the scalar field $\sigma$ is a constant
$\sigma=\sigma_0$ and $V\left( \sigma_0 \right)=0$.
Furthermore as long as we consider the gravitational wave, we do
not consider the fluctuation of the scalar field $\sigma$.
This also shows that the metric $\e^\sigma g_{\mu\nu}$ which appears in
the Lagrangian density of matter $\mathcal{L}_\mathrm{matter}$ is
different from the metric $g_{\mu\nu}$ only by a constant scale
transformation, which effectively gives the change of the gravitational
constant $\kappa^2 \to \kappa^2 \e^{\sigma_0}$.
Thus as long as we neglect the cosmic expansion, the propagation
of the gravitational wave in the $F(R)$ gravity is not qualitatively
changed from that
in the Einstein gravity.

By the variation of the action in (\ref{JGRG23}) with respect to
the metric $g_{\mu\nu}$, we obtain the Einstein equation in the
Einstein frame,
\begin{equation}
\label{EE1}
R_{\mu\nu} - \frac{1}{2} g_{\mu\nu} R =
3 \partial_\mu \sigma \partial_\nu \sigma
+ g_{\mu\nu} \left( - \frac{3}{2} g^{\rho\sigma} \partial_\sigma \phi
\partial_\sigma \sigma
   - V(\sigma) \right)
+ \kappa^2 \e^\sigma T_{\mu\nu}\, .
\end{equation}
Naively if the first two terms are dominant compared with the last term
$\kappa^2 \e^\sigma T_{\mu\nu}$ as in the vacuum,
we can neglect the contribution from the matter.
On the other hand, if the last term $\kappa^2 \e^\sigma T_{\mu\nu}$ is dominant
as in the dense matter, the expansion of the universe, which could be generated
by the first two terms, could be negligible.

In case of the $F(R)$ gravity, there appears a scalar mode corresponding to
$\sigma$.
The mass of the scalar mode should be very small in the bulk but the mass can
become
large inside the matter by the Chameleon mechanism~\cite{Khoury:2003rn}.
Therefore the propagation of the scalar mode is  a little bit complicated.

The propagation of the graviton in other kinds of the modified gravity theories
has been
also actively
investigated, see,
for instance,~\cite{Akrami:2018yjz,Casalino:2018tcd,Granda:2018tzi,
Berti:2018cxi,Akrami:2017cir,Paschalidis:2017qmb,Jana:2017ost}

\section{Conclusions \label{SecV}}

In the present paper, we have analyzed the propagation
of gravitational waves in the medium
in detail. We have shown how the propagation of gravitational waves
could be changed by the medium.

In general, the radiation is made of the quanta or massless particles
at high temperature.
Usually the radiation consists of photons, which are quanta of the
electromagnetic field.
We should note that the radiation-dominated stage of the universe can
be realized not only by the real radiation but by the scalar-tensor theory.
Then we have shown how to distinguish the radiation dominated universe
generated by the real radiation with that generated
by the scalar-tensor tehory.

Motivated with the above observations, we have investigated the propagation of
gravitational waves in the medium.
Especially it has been found that the propagation of gravitational waves in the
thermal radiation
in general relativity is different from that in the scalar-tensor theory.
Furthermore, we have explored the propagation of gravitational waves in
the uniform magnetic field and it has been found that the effects from the
magnetic field to the propagation could not be negligible.
For the small object like magnetar, the perturbation with respect to $\kappa^2 B^2$
could be valid but for the large object of galaxy size like large magnetic field
the perturbation breaks down.
For the large object, the adiabatic approximation gives more reasonable results.
Note that we limited to flat
space background where there is no qualitative difference between General
relativity and say, $F(R)$ gravity.

At the next stage, it would be extremely interesting to extend our study
for evolving cosmological background.
It is known that the evolution of gravitons in accelerating cosmologies
for
the case of extended gravity which has  been considered in
Ref.~\cite{Capozziello:2017vdi} is qualitatively different from that of
General Relativity.
Then, the account of cosmic magnetic field may even increase this
qualitative difference. Then, as
first proposal for future possible extensions of the present
work one can study the gravitational waves propagation
in anisotropic (Bianchi) universe with magnetic fields.
Moreover, it has been examined that the stochastic background of gravitational
waves can
be tuned by the effect of $F(R)$ gravity~\cite{Capozziello:2007zza}.
This again may be generalized for the presence of cosmic magnetic field.

Finally, it is very interesting to mention that
   there is a possibility of the existence of some relations between
primordial
magnetic fields and primordial gravitational fields~\cite{Bamba:2014vda}.
This may be a clue to find a fundamental connection between electromagnetism
and
gravitation, which would be similar to that between thermodynamics and gravity.
From other side, such study may give further bounds to gravitational waves
propagation at the early universe with primordial magnetic fields.

\section*{Acknowledgments}

This work is supported (in part) by
MEXT KAKENHI Grant-in-Aid for Scientific Research on Innovative Areas
``Cosmic Acceleration'' (No. 15H05890) (SN),
by MINECO (Spain), project FIS2016-76363-P (SDO) and JSPSS17116 short-term
fellowship (SDO). In addition, the work of KB is supported by the JSPS KAKENHI
Grant
Number JP 25800136 and Competitive Research Funds for
Fukushima University Faculty (17RI017).

\end{document}